\documentclass[prb,preprint]{revtex4}

\usepackage{amsmath}

\newcommand{\rv}{\mathbf{r}}
\newcommand{\rvhat}{\hat {\mathbf r}}

\usepackage{graphicx}

\begin{document}

\title{Modification of Coulomb's law in closed spaces}
\author{Pouria Pedram}
\email{pouria.pedram@gmail.com}
\affiliation{Plasma Physics Research Center, Science and Research
Campus, Islamic Azad University, Tehran, Iran}


\begin{abstract}
We obtain a modified version of Coulomb's law in two- and
three-dimensional closed spaces. We demonstrate that in a closed
space the total electric charge must be zero. We also discuss the
relation between total charge neutrality of a isotropic and
homogenous universe to whether or not its spatial sector is closed.
\end{abstract}

\maketitle

\section{Introduction}\label{sec1}
\begin{figure}
\centering
\includegraphics[width=8cm]{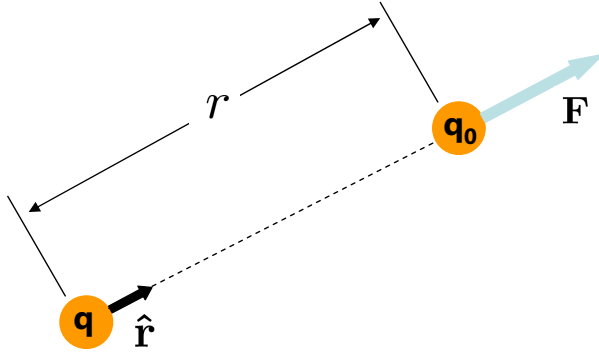}
\caption{The electric force between two point charges separated by a
distance $r$.} \label{fig1}
\end{figure}

One of the fundamental forces of nature is the electromagnetic force
between charged particles. The interaction between charged
particles in flat spaces is governed by Coulomb's law in two and three dimensions\cite{Jackson,Heering}
\begin{align}
\label{coulumb0}
\mathbf{E}=\frac{\mathbf{F}}{
q_0}=\frac{q}{2\pi \epsilon_0 r} \hat \rv &&&\mbox{(two dimensions)} \\
\mathbf{E}=\frac{q}{4\pi \epsilon_0 r^2} \hat \rv &&& \mbox{(three dimensions)}. \label{coulumb}
\end{align}
Here, $\mathbf{E}$ is the electric force per unit charge and $\hat
\rv$ is the unit vector directed from the first charge $q$ to  the
second charge $q_0$ (see Fig.~\ref{fig1}).

There are two basic assumptions that are valid in flat spaces for
electrostatic forces. One is the superposition principle, which
states that at any point in space, the total electric field of a
group of charges equals the vector sum of the electric fields due to
the individual charges. The other assumption in flat space is the
absence of any restriction on the number of positive and negative
charges; that is, we may have an arbitrary number of positive and
negative charges with zero or nonzero total value. As we shall see,
in curved spaces the correctness of the former is not clear and the
latter assumption is not valid.\cite{Landau}

The first attempt to solve electrostatic
problems in curved space were done by Enrico Fermi in
1921.\cite{Fermi} In this paper Fermi discussed the correction to
the electric field of a single point charge held at rest within a
gravitational field to first order in the gravitational
acceleration. A few years later, Edmund T.\ Whittaker solved this problem exactly
both in the homogeneous gravitational field and in the Schwarzschild
geometry cases.\cite{Whittaker} These efforts were further developed
by E.\ T.\ Copson.\cite{Copson}

In this paper we obtain the modified form of Coulomb's law in two-
and three-dimensional closed curved spaces. We show that in the
vicinity of charges where the effect of curvature is negligible, we
recover Coulomb's law. The geometrical interpretation of this result
is that we can always find a tangent flat space for any point on a
curved space. We also demonstrate the failure of the second
assumption in these spaces. Finally, we discuss the charge
neutrality of a closed isotropic and homogenous universe.

\section{Mathematical preliminaries}
Because Coulomb's law is valid only in flat spaces, we outline some
important properties of these spaces. In a flat space there always
exists a (Cartesian) coordinate system where the distance between
two infinitesimally close points can be written as
\begin{equation}\label{metric1}
ds^2=\sum_{i=1}^{d}(dx^i)^2,
\end{equation}
where $x^i$ are the coordinates, $d$ is the dimension of the space,
and $ds^2$ is the line element. We can define the metric of the
space as a second-order covariant tensor $g_{ij}$, namely
\begin{equation}\label{metric2}
ds^2=\sum_{i,j}g_{ij}dx^idx^j.
\end{equation}
Thus, in flat spaces, it is always possible to find a coordinate
system in which the metric is diagonal with constant elements
\begin{equation}\label{metric3}
g_{ij}=\pm \delta_{ij},
\end{equation}
where $\delta_{ij}$ is the kronecker delta defined as
\begin{equation}\label{kronecker}
\delta_{ij}=
\begin{cases}
1 & i=j,\\
0 & i\ne j.
\end{cases}
\end{equation}
In particular, Euclidean space corresponds to
$g_{11}=g_{22}=g_{33}=g_{44}=1$, and Lorentzian space
corresponds to $g_{11}=g_{22}=g_{33}=1,g_{44}=-1$. A
well-known example is the three-dimensional spherical coordinate
system specified by $r$, $\theta$, $\phi$, and the line element
\begin{equation}\label{polar1}
ds^2=dr^2+r^2d\theta^2+r^2\sin^2\theta d\phi^2,
\end{equation}
or its equivalent metric tensor in matrix form
\begin{equation}\label{polar2}
\left[g_{ij}\right]=
\begin{bmatrix}
1 & 0 & 0 \\ 0 & r^2 & 0 \\
0& 0 & r^2\sin^2\theta \\
\end{bmatrix}.
\end{equation}
Using a suitable choice of the coordinate transformation
\begin{equation}
\begin{array}{l}
z=r \cos\theta,\\
x=r\sin\theta \cos\phi,\\
y=r\sin\theta\sin\phi,
\end{array}
\end{equation}
we obtain the diagonal form of the line element
\begin{equation}
ds^2=dx^2+dy^2+dz^2,
\end{equation}
and the metric tensor
\begin{equation}
\left[g_{ij}\right]=
\begin{bmatrix}
1 & 0 & 0 \\ 0 & 1 & 0 \\
0& 0 & 1 \\
\end{bmatrix}.
\end{equation}

We now consider the surface of a 2-sphere as a two-dimensional
closed curved space (see Fig.~\ref{fig2}). The distance between two
infinitesimally close points on this hypersurface is
\begin{equation}\label{metric-2sphere}
ds^2=R^2d\theta^2+R^2\sin^2\theta d\phi^2,
\end{equation}
where $R$ is the radius of the sphere, which takes a constant value.
For this case it is not possible to find a proper coordinate
transformation to diagonalize the metric tensor with constant
elements. In other words, we cannot find a new set of variables
$u(\theta,\phi)$ and $v(\theta,\phi)$ so that the line element takes
the simple form $ds^2 = du^2+dv^2$.

To be more precise, let us review the tensorial properties of the
2-sphere. We will show that because all components of the curvature tensor
are nonzero, the metric tensor of a 2-sphere is not
Euclidean.\cite{Dalarsson} The covariant metric tensor of the
surface of a sphere can be written in the matrix form
(\ref{metric-2sphere})
\begin{equation}\label{metricc}
\left[g_{ij}\right]=R^2
\begin{bmatrix}
1 & 0 \\ 0 & \sin^2\theta \\
\end{bmatrix},
\end{equation}
where $i,j$ are elements of $\{\theta,\phi\}$. We have the
following form for the contravariant metric tensor
\begin{equation}
\left[g^{ij}\right]=R^{-2}
\begin{bmatrix}1 & 0 \\ 0 & (\sin\theta)^{-2} \\
\end{bmatrix}.
\end{equation}
The Christoffel symbols of the first kind are defined as
\cite{Dalarsson}
\begin{equation}\label{Christoffel1}
\Gamma_{i,jk}=\frac{1}{2}(\partial_kg_{ij}+\partial_jg_{ki}-\partial_ig_{jk}).
\end{equation}
By substituting Eq.~(\ref{metricc}) into Eq.~(\ref{Christoffel1}) we have
$\Gamma_{\theta,\theta \theta}=0$, $
\Gamma_{\theta,\theta \phi}=\Gamma_{\theta,\phi \theta }=0$, $
\Gamma_{\theta,\phi\phi}=-R^2\sin\theta \cos\theta$,
$\Gamma_{\phi,\theta \theta}=0$,
$\Gamma_{\phi,\theta \phi}=\Gamma_{\phi,\phi \theta }=+R^2\sin\theta \cos\theta$, and
$\Gamma_{\phi,\phi \phi}=0$.
The Christoffel symbols of the second kind are also defined as
\begin{equation}
\Gamma^i_{jk}=g^{il}\Gamma_{l,jk}.
\end{equation}
Thus, we have
$\Gamma^{\theta}_{\theta \theta}=0$,
$\Gamma^{\theta}_{\theta \phi}=\Gamma^{\theta}_{\phi \theta }=0$,
$\Gamma^{\theta}_{\phi\phi}=-\sin\theta \cos\theta$,
$\Gamma^{\phi}_{\theta \theta}=0$,
$\Gamma^{\phi}_{\theta \phi}=\Gamma^{\phi}_{\phi \theta }=+(\sin\theta)^{-1} \cos\theta$,
and $\Gamma^{\phi}_{\phi \phi}=0$.

We can use the Christoffel symbols to define the Riemann
curvature tensor
\begin{equation}\label{Riemann}
{\cal R}^{i}_{jkl}=\partial_{k}\Gamma^i_{lj}-\partial_l\Gamma^i_{kj}+\Gamma^m_{lj}\Gamma^i_{km}-\Gamma^m_{jk}\Gamma^i_{lm}.
\end{equation}
The Riemann curvature tensor in $d$-dimensional space has $d^4$
components. These components are not all independent, and the number
of independent components are given by
\begin{equation}
n=\frac{d^2(d^2-1)}{12},
\end{equation}
which is equal to $1$ for a 2-sphere ($d=2$). We can also define the
Ricci tensor
\begin{equation}
{\cal R}_{ij}= {\cal R}^{k}_{ijk}=g^{kl}{\cal R}_{lijk},
\end{equation}
and the Ricci scalar
\begin{equation}\label{Ricci}
{\cal R}=g^{ij}{\cal R}_{ij}=g^{ij}g^{lk}{\cal R}_{kijl},
\end{equation}
where the former is a symmetric tensor and the latter is proportional
to the Gauss curvature. The only independent component of the
Riemann curvature tensor for our case is ${\cal R}_{\theta
\phi\theta \phi}$, which by Eq.~(\ref{Riemann}) is
\begin{equation}
{\cal R}_{\theta \phi\theta
\phi}=-\frac{1}{2}\partial_{\theta}\partial_{\theta}g_{\phi\phi}+g_{\phi\phi}(\Gamma^{\phi}_{\theta\phi})^2=(R\sin\theta)^2.
\end{equation}
Thus, we can obtain the Gauss curvature of the 2-sphere given by
Eq.~(\ref{Ricci}):
\begin{subequations}
\begin{align}
\frac{{\cal R}}{2}&= g^{\theta\theta}g^{\phi\phi}{\cal R}_{\theta
\phi\theta \phi}-g^{\theta\phi}g^{\theta\phi}{\cal R}_{\theta
\phi\theta \phi}
=(g^{\theta\theta}g^{\phi\phi}-g^{\theta\phi}g^{\theta\phi}){\cal
R}_{\theta \phi\theta \phi},\\ &= |g^{ij}|{\cal R}_{\theta
\phi\theta \phi}=\frac{{\cal R}_{\theta \phi\theta
\phi}}{g}=\frac{1}{R^2},
\end{align}
\end{subequations}
which is equal to the square inverse of its radius.

\section{Electric field on a two-dimensional spherical space}
\begin{figure}
\centering
\includegraphics[width=8cm]{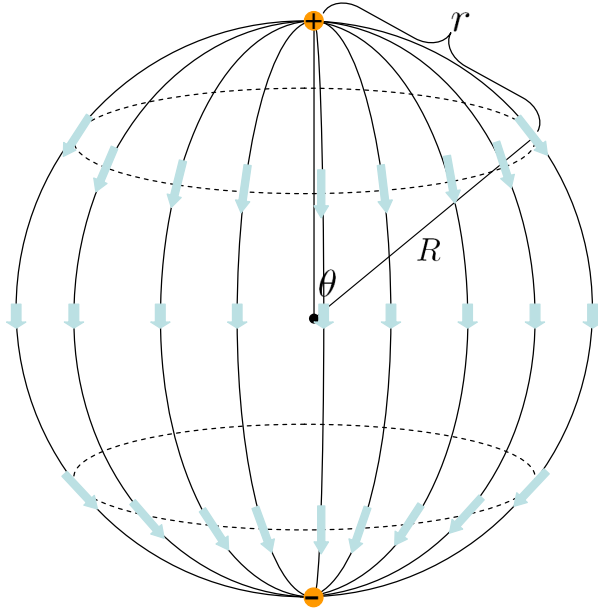}
\caption{The electric field on a 2-sphere in the presence of
positive and a negative point charges located at the north and south
poles, respectively.} \label{fig2}
\end{figure}

We consider a 2-sphere as a simple two-dimensional closed curved
space (see Fig.~\ref{fig2}). The points on this space satisfy
\begin{equation}
x^2+y^2+z^2=R^2.
\end{equation}
Now, put a positive point charge $q$ at its north pole. We assume
that the electric fields exist only on the sphere's surface. In this
situation, the field's lines come out from the north pole and after
passing the equator meet each other at the south pole. The
intersection point of the ingoing field lines corresponds to the
presence of a negative charge. Thus, we will observe a negative
point charge $-q$ at the south pole of the sphere. The presence of
the negative charge shows that there is a one to one correspondence
between positive and negative charges in this space. Therefore, the
total charge on the sphere will be zero
\begin{equation}
\sum_i q_i=0.
\end{equation}
To obtain the electric field, we use Gauss's law in
2 dimensions
\begin{equation}
\oint \mathbf{E} \cdot d\mathbf{S}=\frac{q}{\epsilon_0},\label{Gauss}
\end{equation}
where the integration is over a one-dimensional closed curve. For
instance consider the upper dashed line in Fig.~\ref{fig2} as the
integration contour. Because this path encloses both $q$ and $-q$,
it is not possible to relate the electric field to one charge only.
So the resulting electric field can be decomposed into two
components which are related to each charge. For this case the
integration contour is a circle with circumference $2\pi
R\sin\theta$ (see Fig.~\ref{fig2}). The electric field on the sphere
is
\begin{equation}
\mathbf{E}=\frac{q}{2\pi \epsilon_0 R
\sin\theta} \hat{\boldsymbol\theta},
\end{equation}
where $\theta$ is the usual polar coordinate and $R$ is the radius
of the sphere. If we define $r$ as the distance from the north pole
on the sphere, we can rewrite the electric field in terms
of $r$:
\begin{equation}
\label{thiseq}
\mathbf{E}=\frac{q}{2\pi \epsilon_0 R \sin (r/R)}\hat \rv.
\end{equation}
For a large value of $R$ or in the vicinity of the positive charge
($r\ll R$), Eq.~\eqref{thiseq} reduces to the flat
two-dimensional form of Eq.~(\ref{coulumb0})
\begin{equation}
\mathbf{E}\simeq\frac{q}{2\pi \epsilon_0 r} \hat \rv.
\end{equation}
Moreover, because the negative charge $-q$ is located at
infinity in this limit, its effect is negligible in
the vicinity of the positive charge.

Note that, because the integration contour contains both positive
and negative charges, it may seem that the superposition law fails
in curved spaces. But, because Maxwell's equations are linear in any
spacetime, this conclusion is not true. The reason why the
superposition principle seems to fail is related to the topology of
the space: A general solution to Maxwell's equation will not be
single-valued in a closed space, and additional conditions must be
imposed to ensure that the electric field is globally well-defined.
These additional conditions require the existence of an additional
charge and seem to imply a failure of the superposition principle.
However, the superposition principle is not violated, because the
correct solution is a sum $\mathbf{E}_1 + \mathbf{E}_2 +
\mathbf{E}_3$, where $\mathbf{E}_1$ is the field of the positive
charge, $\mathbf{E}_2$ is the field of the negative charge, and
$\mathbf{E}_3$ is a solution to the homogeneous equation (Laplace's
equation) that must be added to account for the topological
conditions.

\section{Electric field on a three-dimensional spherical space}
\begin{figure}
\centerline{\begin{tabular}{ccc}
\includegraphics[width=6cm]{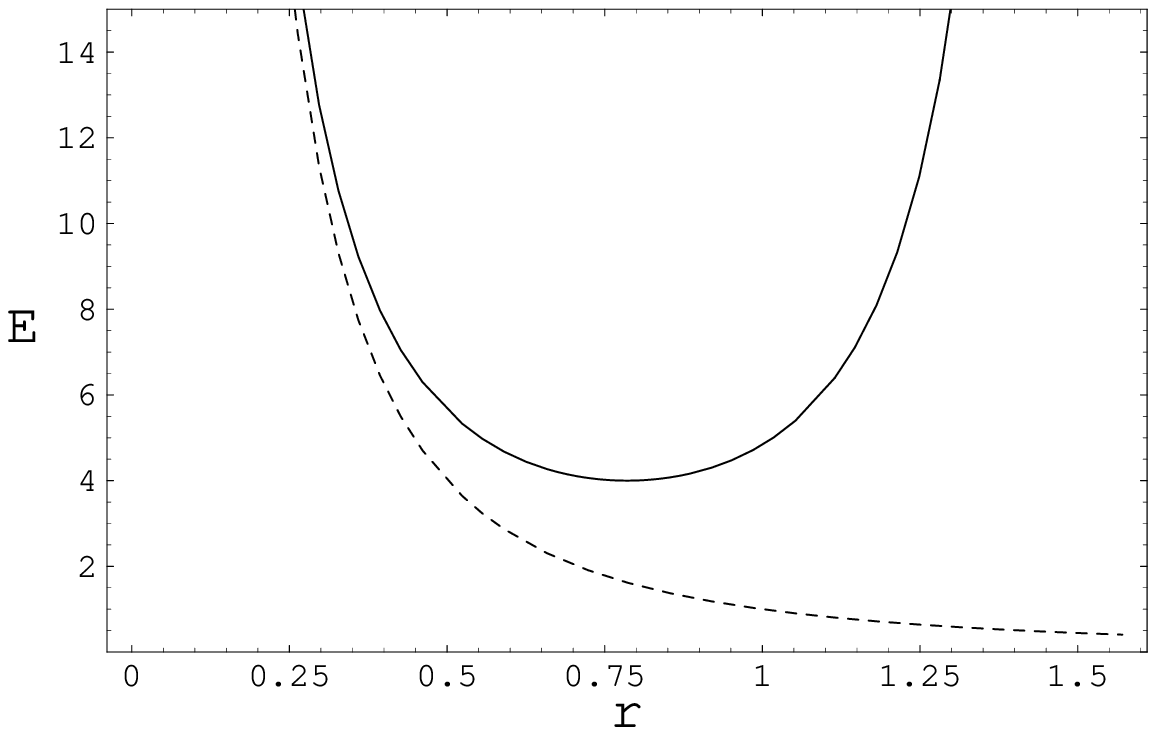}&
\includegraphics[width=6cm]{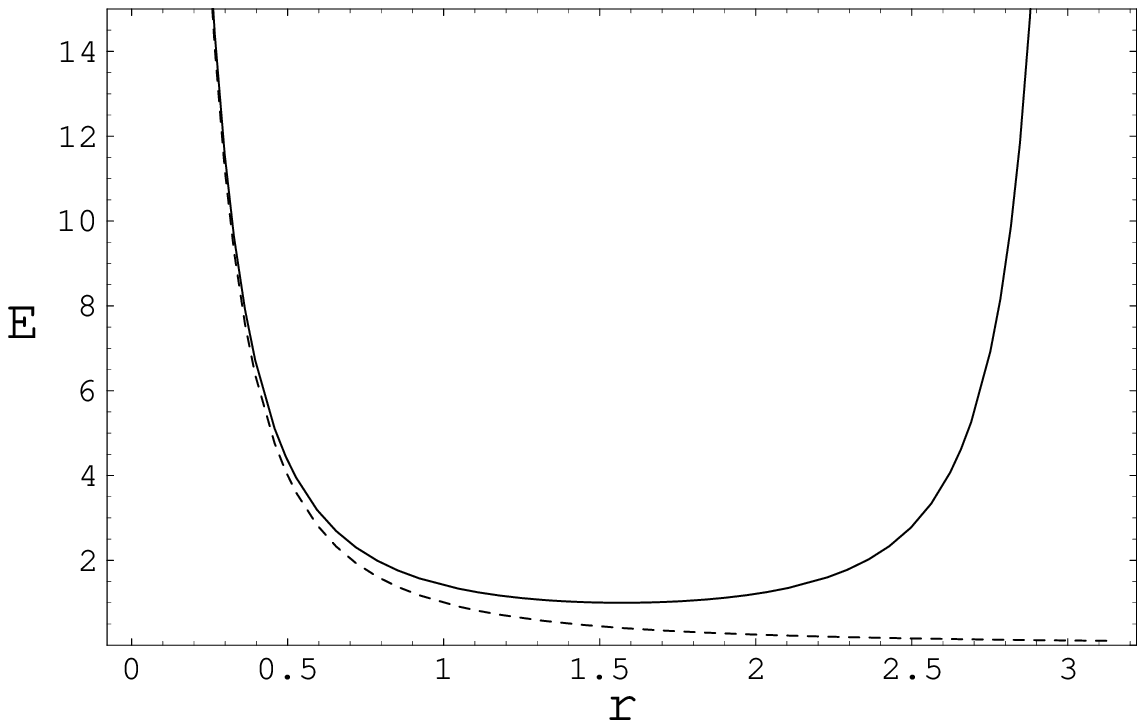}&
\includegraphics[width=6cm]{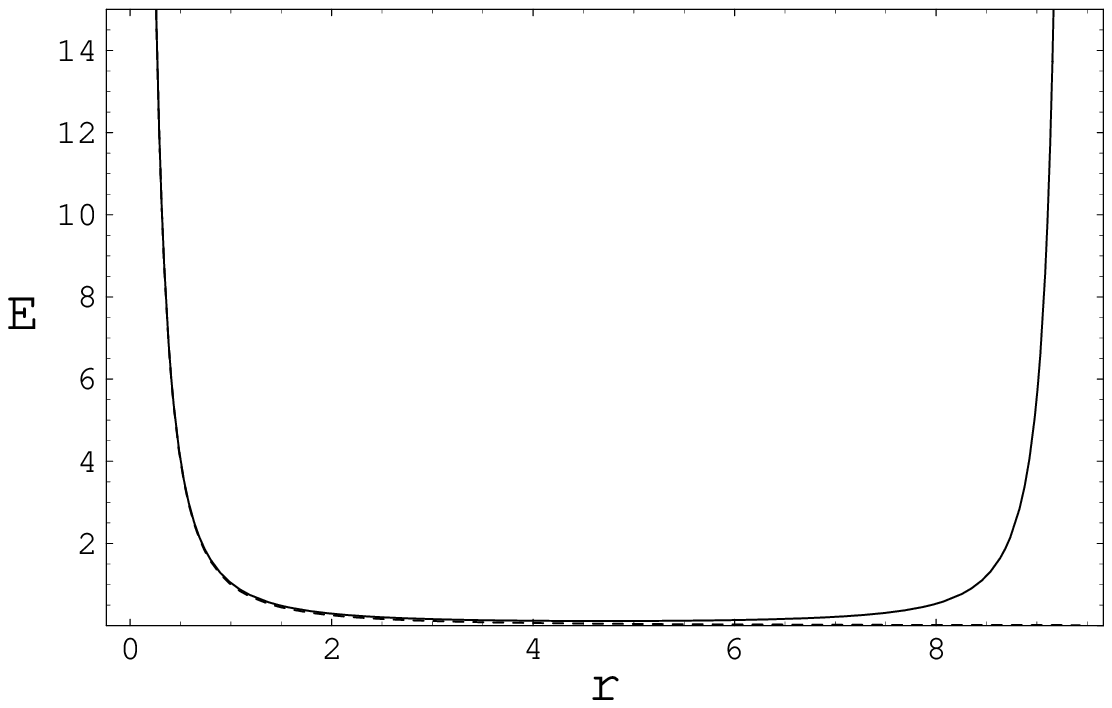}
\end{tabular}}
\caption{The electric field on a 3-sphere, Eq.~(\ref{electric-C})
(solid line), and Coulomb's law, Eq.~(\ref{coulumb}) (dashed line),
for (a) $R=1/2$, (b) $R=1$, and (c) $R=3$ with
$q/(4\pi\epsilon_0)=1$.} \label{fig3}
\end{figure}

To have a more realistic model, we consider a 3-sphere
which is a set of points equidistant from a fixed central point in
four-dimensional Euclidean space.\cite{Peterson,Henderson} The points
on this three-dimensional hypersurface satisfy the relation
\begin{equation}\label{3phere}
x^2+y^2+z^2+\tau^2=R^2,
\end{equation}
and can be expressed by
$z=R \sin \chi \cos\theta$,
$x=R \sin \chi \sin\theta \cos\phi$, $y=R \sin \chi \sin\theta
\sin\phi$, $R$ is the radius of the
3-sphere, $\chi$ is the extra polar coordinate, and $\tau=R \cos \chi$,
is the 4th coordinate. Now, consider a
positive charge located at the north pole of the 3-sphere. We require
that the electric field is confined to the three-dimensional
hypersurface. Thus, the electric field cannot propagate outside the
3-sphere. By using the constraint equation (\ref{3phere}),
we can rewrite the line element of this space as
\begin{equation}\label{metric-3sphere}
ds^2=dx^2+dy^2+dz^2+\frac{(xdx+ydy+zdz)^2}{R^2-(x^2+y^2+z^2)}.
\end{equation}
If we use the corresponding spherical coordinates $\chi$,
$\theta$, $\phi$ instead of the coordinates $x$, $y$, $z$, the line
element takes the form
\begin{equation}
ds^2=R^2\left[d\chi^2+\sin^2\chi(d\theta^2+\sin^2\theta\,
d\phi^2)\right].
\end{equation}
To obtain the electric field, we need to use Gauss's law
(\ref{Gauss}) over a closed two-dimensional surface with constant $r$
(or $\chi$) with the line element
\begin{equation}
ds^2=R^2\sin^2\chi(d\theta^2+\sin^2\theta\, d\phi^2),
\end{equation}
which is similar to the case of the 2-sphere ($R\rightarrow
R\sin\chi$). The area of this hypersurface is equal to
$\mathbf{S}=4\pi R^2 \sin^2\chi$. The use of Gauss's law
results in the following form for the electric field
\begin{equation}\label{electric-C}
\mathbf{E}=\frac{q}{4\pi \epsilon_0 R^2 \sin^2\chi}\hat
{\boldsymbol \chi},
\end{equation}
Equation~\eqref{electric-C} reduces to Coulomb's law (\ref{coulumb})
in the vicinity of the north pole ($r\ll R$). To show this result, we
rewrite the line element (\ref{metric-3sphere}) in terms of $r'$,
$\theta$, $\phi$
\begin{align}
ds^2&=\frac{dr'^2}{1-r'^2/
R^2}+r'^2(d\theta^2+\sin^2\theta d\phi^2),\\&=R^2\left[
\frac{dr''^2}{1-r''^2}+r''^2(d\theta^2+\sin^2\theta d\phi^2)
\right],\label{frw}
\end{align}
where $r'^2=x^2+y^2+z^2$ and $r''=
r'/R$. Moreover, we can obtain the relation between
$r'$ and the radius of a sphere on this hypersurface
\begin{equation}
r=\!\int_{0}^{r'}\frac{dr'}{\sqrt{1-r'^2/R^2}}=R\sin^{-1}r'/R=R\chi,
\end{equation}
which results in the electric field
\begin{equation}
\mathbf{E}=\frac{q}{4\pi \epsilon_0 R^2 \sin^2(\frac{
r}{ R})} \rvhat,
\end{equation}
or
\begin{equation}
\mathbf{E}\simeq\frac{q}{4\pi \epsilon_0 r^2} \rvhat,
\end{equation}
in the vicinity of the positive charge ($r\ll R$).
In other words, we can recover the usual form of the electrostatic
forces in the regions where the effect of the curvature is
negligible. This result also shows that, similar to the
two-dimensional case, the total charge of the 3-sphere should be zero.
In Fig.~\ref{fig3} we have plotted Coulomb's law and its
modified version on the 3-sphere for various values of $R$. As it
can be seen, the correspondence between the two
increases as $R$ increases.

In reality, we live in a four-dimensional curved space-time with a
matter distribution on it. Following the theory of general
relativity, the curvature of the universe comes from its mass and
energy. Any massive object distorts its surrounding space-time. If
we assume that the universe is homogenous and isotropic, space-time
can be expressed by the Friedmann-Robertson-Walker metric as
\cite{Landau2,1,Harvey}
\begin{equation}\label{metric}
ds^2=-dt^2+R^2(t) \left[\frac{ dr''^2}{
1-k\,r''^2}+r''^2(d\theta^2+\sin^2\theta\,d\phi^2) \right],
\end{equation}
where $R(t)$ is the scale factor, and $k=+1,0,-1$ corresponds to a
closed, flat, or open universe, respectively. For a closed universe
($k=1$), the spatial section of the metric has the form
Eq.~(\ref{frw}), and thus can be considered as a 3-sphere imbedded
in four-dimensional space-time, where the scale factor $R(t)$ plays
the role of its radius.\cite{Landau2,Peterson} Observations of our
universe suggest that the spatial section of the universe is flat at
the present time ($k=0$). What would happen if we had a universe
with positive curvature ($k=1$)? In early times, in the interval
between a millisecond to a second after the big bang,\cite{Phillips}
the radius of the universe was very small and the effect of the
curvature was considerable. At the end of the radiation era, when
the temperature of the universe decreased, charged particles could
be created. For a closed universe the curvature would allow the
charged particles to be created in pairs. Consequently, the universe
at large scales would be neutral. Thus, a closed universe results in
the charge neutrality of the universe, but the inverse statement is
not necessarily true. Hence, although observations show that our
universe is neutral at large scales, the flatness of our universe
indicates that the neutrality is only a coincidence from this point
of view.

\begin{acknowledgments}
The author wishes to thank the referee for constructive comments and
suggestions.
\end{acknowledgments}

\end{document}